# Generalized Super-Cerenkov Radiations in Nuclear and Hadronic Media

D. B. Ion[1,2)]

[1)] IFIN-HH, Bucharest, P.O.Box MG-6, Magurele, Romania
[2)] TH-Division, CERN, CH-1211 Geneva 23, Switzerland

**Abstract**: Generalized Super-Cerenkov Radiations (SCR), as well as their SCR-signatures are investigated and classified. Two general SCR- coherence conditions are found as two natural extremes of the same spontaneous particles decay in (dielectric, nuclear or hadronic) media. The main results on the quantum theory of the SCR-phenomena as well as the results of the first experimental test of the super-coherence conditions, obtained by using the experimental data from BNL are presented. The new concepts such as: SCR-gluons, SCR-W-bosons and SCR-Z-bosons, all three suggested by elementary particle classification, are introduced. The gluonic Super-Cerenkov-like radiation, first introduced here, is schematically described. The interpretation of some recent RHIC results as signature of the SCR-gluons is suggested.

## 1. Introduction

The classical theory of the radiation emitted by charged particles moving with superluminal velocities were traced back to Heaviside[1], Des Coudres [2] and Sommerfeld. [3] [see the papers [1-9]: T. R. Kaiser [10] , Jelley [11], A.A. Tyapkin [12] , B.B. Govorkov [13]]. In fact, Heaviside considered the Cerenkov radiation in a nondispersive medium. He considered this topic many times over the next 20 years, deriving most of the formalism of what is now called Cerenkov radiation and which is applied in the particle detectors technics (e.g., RICH-detectors).

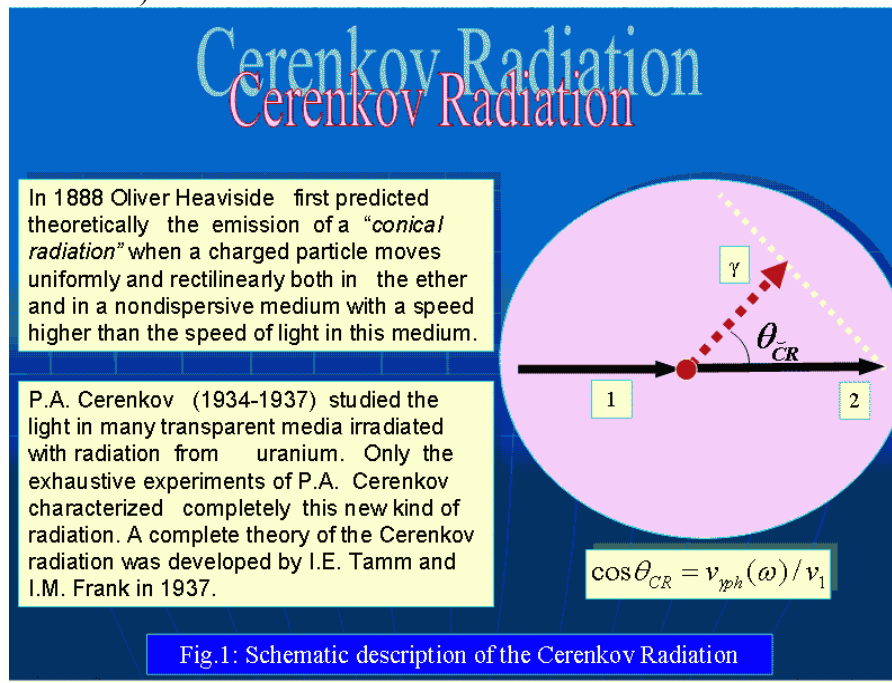

Fig.1: Schematic description of the Cerenkov Radiation

So, doing justice (see again the papers [10-13]) to Heaviside and Sommerfeld, we must recall that the complete classical theory of the CR phenomenon in a dispersive medium was first formulated by Frank and Tamm in 1937. This theory explained all the main features of the radiation observed experimentally by Cerenkov (see Fig. 1) .
The quantum theoretical approach to the CR-problem was developed by many authors [see V. L. Ginsburg, [7] and references in the books [8,9]] .
In essence, it was revealed by the Heaviside, Cerenkov, Tamm and Frank that a charged particle moving in a transparent medium with an refractive index, $n_\gamma$, and having a speed $v_1$ greater than phase velocity of light $v_{\gamma ph} = c/n = n_\gamma^{-1}$ will emit electromagnetic radiation, called *Cerenkov radiation* (CR), at an polar emission angle $\theta_{CR}$ relative to the direction of motion given by the relation (see Fig. 1).:

$$\cos\theta_{CR} = \frac{v_{\gamma ph}(\omega)}{v_1} \quad (1)$$

In this paper we adopted the usual system of units from particle physics ($\hbar = c = 1$)
The remarkable properties of the Cerenkov radiation find wide applications in practice especially in high energy physics where it is extensively used in experiments for counting and identifying relativistic particles [via Ring Imaging Cherenkov (RICH)-Detectors] in the fields of elementary particles, nuclear physics and astrophysics.

**2. Mesonic Cerenkov-like effects in hadronic media**.

The idea that meson production in nuclear interactions may be described as a process similar to the Cerenkov radiation has considered by Wada [14], Ivanenko [15] Blohintev si Indenbom [16], Czyz, Ericson, Glashow [18], Czyz Glashow [19], Smrz [20] and D. B. Ion.[21,22], D.B.Ion and F. Nichitiu [23], Zaretski and Lomonosov [24], Dremin [25,26]. In 1971, in the Doctoral Thesis [21], it was developed a general classical and quantum theory of the mesonic Čerenkov-like radiation in hadronic media. Moreover, the vector-mesonic Cerenkov-like radiation as well as baryonic Cerenkov-like effects in nuclear and hadronic media was also introduced for the first time in Ref. [21]. Then, it was predicted completely the properties of the mesonic Čerenkov-like radiation in the case when the mesonic refractive index is given by a single pole approximation and obtained a good agreement with the integrated cross section of the single meson production in the hadronic collisions (see Ref. [21-23]).
In 1990-19995, we have extended [27-36] these ideas to the nuclear media where gamma Cherenkov radiation (NGCR) and pion Cerenkov-like radiations (NMCR) should be possible to be emitted from charged particles moving through nuclei with a velocity larger than the phase velocity of photons or/and pions in the nuclear media. The refractive indices of the gamma ($n_\gamma$), meson ($n_M$), nucleon ($n_N$), was calculated by using Foldy-Lax formula (see formula (1) in Ref. [35]) and the experimental pion-nucleon cross sections combined with the dispersion relations predictions, the refractive index of pions in the nuclear media has been calculated [31-36]. Then, the detailed predictions for the spontaneous pion emission as nuclear pionic Cherenkov radiation (NPICR) inside the nuclear medium are obtained and published in Refs. [32, 34-37]. The following essential characteristic features of the Cherenkov pions are predicted.
(i) The true coherent pion emission as *nuclear pionic Cherenkov-like radiation* (NPICR) is possible in the following three energy bands:
CB1-NPICR band for: $190 MeV \leq \omega \leq 315 MeV$ for all $\pi^{\pm,0}$
CB2-NPICR band for: $910 MeV \leq \omega \leq 960 MeV$ only for $\pi^+$, and

CB3-NPICR band for : $80 GeV \leq \omega \leq 1000 GeV$ for all $\pi^{\pm,0}$
in the nuclear reactions such as: $^{208}PbN \Rightarrow {}^{208}PbN\pi$

(ii) For the CB1 the NPICR-differential cross section are peaked at the energy $\omega_m = 260$ MeV for CB1 band and $\omega_m = 930$ MeV for CB2 band when the absorption is neglected and the peak position is shifted up to $\omega_m = 240$ MeV for CB1-band when the absorption is taken into account.

(iii) The NPICR-pions must be coplanar with the incoming and outgoing projectile possessing a strong correlation between the angle of emission $(\theta, \omega_m)$ and the pion ($\omega$) and projectile ($T_p$) energies.

(iv) The energy dependence of the NPICR-peak position is as $E_N^{-2}$.

(v) The A-dependence of the NPICR-peak is near to $A^{-1}$ when the absorbtion is neglected.

The detailed predictions for the spontaneous pion emission as nuclear pionic Cherenkov radiation (NPICR) inside the nuclear medium are obtained and published in Refs. [31]-[35]. Now, it is important to note that in 1999, G.L.Gogiberidze, E.K. Sarkisyan and L.K. Gelovani [37-40] performed the first experimental test of the pionic Cerenkov-like effect (NPICR) in Mg-Mg collisions at 4.3 GeV/c/nucleon by processing the pictures from 2m Streamer Chamber SKM-200. So, after processing a total of 14218 events, which were found to meet the centrality criterion, the following experimental results are obtained:

- The energy distribution of emitted pions in the central collisions have a significant peak (4.1 standard deviations over the inclusive background) (see Fig. 2a)
- The value of the peak energy and its width are

$$E_m^* = [238 \pm 3(stat) \pm 8(syst)] MeV$$
$$\Gamma_\pi = [18 \pm 3(stat) \pm 5(syst)] MeV$$

So, they obtained a good agreement with the position and width of the first pionic Cerenkov-like band predicted by D.B.Ion and W. Stocker in ref. [35].

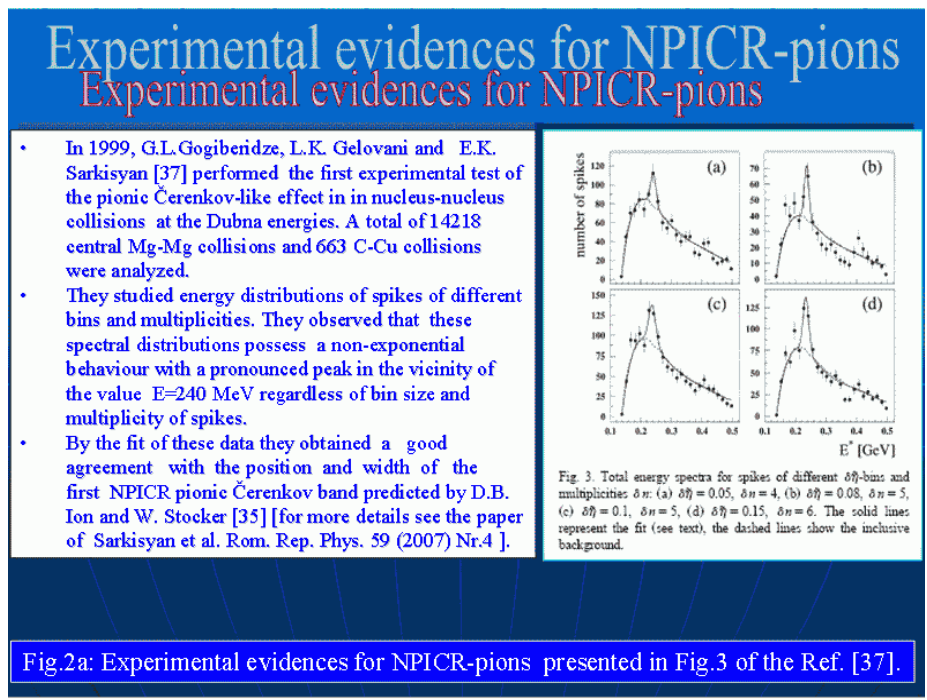

Fig.2a: Experimental evidences for NPICR-pions presented in Fig.3 of the Ref. [37].

The classical variant of the theory of the mesonic Cerenkov-like radiation in hadronic media [21] was applied to the study of single meson production in the hadron-hadron interactions at high energy. This variant is based on on the usual assumption that hadrons are composed from a central core (in which the hadron mass is concentrated surrounded by a large and more diffuse mesonic cloud. (hadronic medium). Then it was shown [21,41,42,43,44,45,46,47,48] that a *hadronic mesonic Cerenkov-like radiation* (HMCR) mechanism (see [21],[41],[42,[48]]), with an mesonic refractive index in hadronic medium given by pole approximation, is able to describe with high accuracy the integrated cross section of the single meson production in the hadron-hadron interactions.

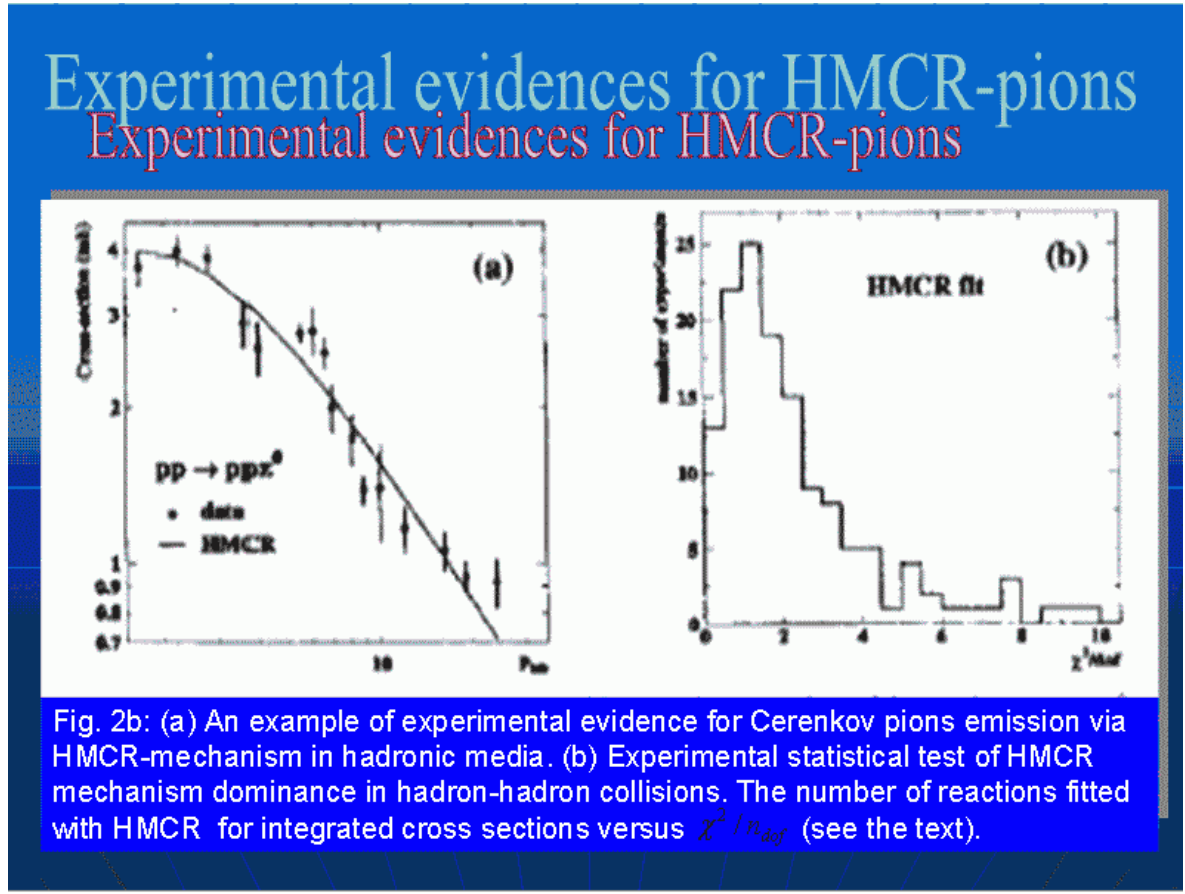

Fig. 2b: (a) An example of experimental evidence for Cerenkov pions emission via HMCR-mechanism in hadronic media. (b) Experimental statistical test of HMCR mechanism dominance in hadron-hadron collisions. The number of reactions fitted with HMCR for integrated cross sections versus $\chi^2/n_{dof}$ (see the text).

To illustrate these important results in Fig. 2b we presentented the measured integrated cross section for the process: $pp \to pp\pi^0$ compared [21,41;42,43,48] with the prediction of mesonic Cherenkov-like radiation (HMCR-mechanism). This result was very encouraging for the extension of the Cerenkov-pions analysis (HMCR-variant) to all processes of single meson productin in hadron-hadron interaction. The results of such analyses are presented in our papers [44,45,46]. Collecting $\chi^2/dof$ for all 139 reactions fitted with the HMCR approach [41,42,44,45,46,47] we obtained the surprisingly good [48] description presented in Fig. 2b.

## 3. Super-Cerenkov Radiation (SCR)

Recent experimental observations of the subthreshold and anomalous Cerenkov radiations (see Fig. 3a) as well as multi-ring phenomena (see Fig. 3b) it was clarified that some fundamental aspects of the CR-theory can be considered as being still open and that more theoretical and experimental investigations on the CR are needed.

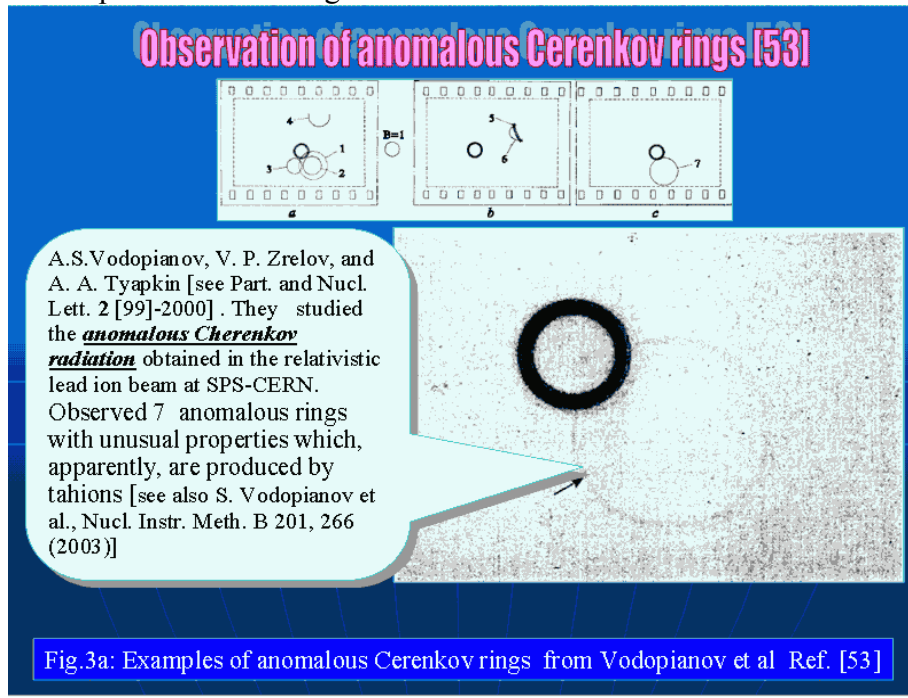

Fig.3a: Examples of anomalous Cerenkov rings from Vodopianov et al Ref. [53]

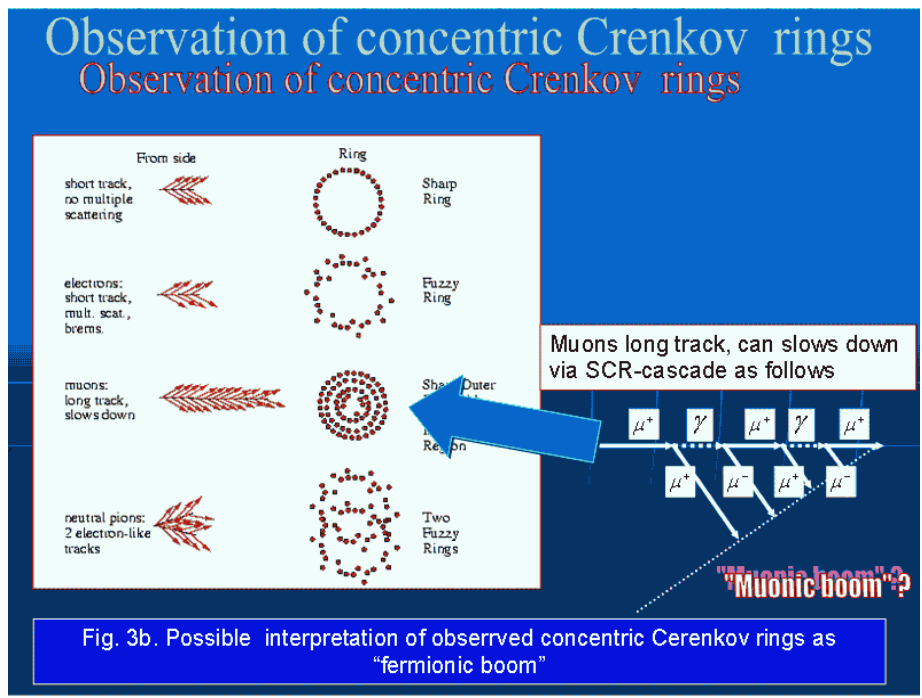

Fig. 3b. Possible interpretation of obserrved concentric Cerenkov rings as "fermionic boom"

Then, theoretical investigations using the CR correct kinematics lead us the discovery that Cerenkov radiation is in fact only the low energy component of a more general phenomenon called by us the Super-Cerenkov radiation (SCR) characterized by the Super-Cerenkov (SCR)-decay condition:

$$\cos\theta_{SC} = v_{\gamma ph} \cdot v_{2ph} \leq 1 \qquad (2)$$

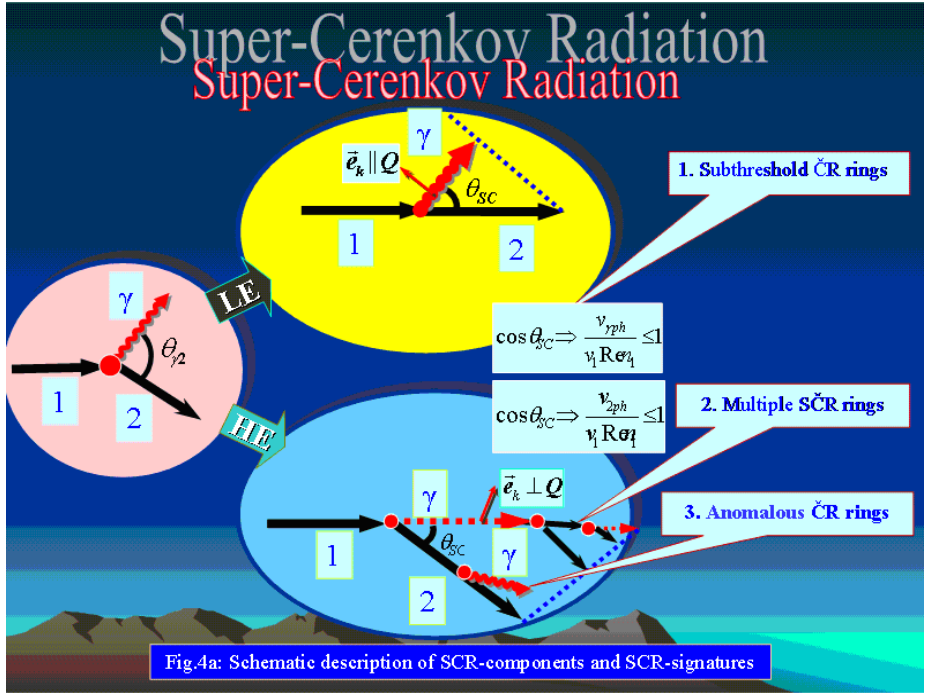

Fig.4a: Schematic description of SCR-components and SCR-signatures

Ideed, let $[E_1,\vec{p}_1,\mathrm{Re}\,n_1(E_1)], [E_2,\vec{p}_2,\mathrm{Re}\,n_2(E_2)], [\omega,\vec{k},\mathrm{Re}\,n_\gamma(\omega)]$ be the energies, momenta, and the refractive indices of particles in given medium from the two-body decay in medium described schematically in Fig.4. Then, using energy-momentum conservation law

$$E_1 = E_2 + \omega, \quad \vec{p}_1 = \vec{p}_2 + \vec{k} \qquad (3)$$

we obtain

$$\cos\theta_{1\gamma} = \frac{\omega}{k}\frac{E_1}{p_1} + \frac{D_2 - D_1 - D_\gamma}{2kp_1},$$
$$D_i \equiv E_i^2 - p_i^2, \quad D_\gamma \equiv \omega^2 - k^2 \qquad (4)$$

Therefore, identifying the phase velocities in medium as

$$v_{\gamma ph} = \frac{\omega}{k}, \quad v_{1ph} = \frac{E_1}{p_1}, \quad v_{2ph} = \frac{E_2}{p_2}, \qquad (5)$$

and considering the second term in Eq.(4) as a small quantum correction from (4) we obtain the condition (2). Hence, the SCR-condition (2) is obtained in a natural way from the energy-momentun conservation law when the influence of medium on the propagation properties of the charged particle is also taken into account and when the quantum corrections are neglected. We note that the refractive is well described by Foldy-Lax formula (see Eq. (11) ) The signature of the SCR-effects are schematically described in Fig.4a.

(i) **Low $\gamma$ – energy SCR-Cerenkov photons** with the usual polarization properties in the decay plane Q, but with an emission angle given by relation

$$\cos\theta_{1\gamma} = v_{1ph}.v_{\gamma ph} \cong \frac{v_{\gamma ph}}{v_1\,\mathrm{Re}\,n_1} \leq 1, \qquad (6)$$

and a shifted threshold described according to formula

$$v_1^{thr} = \frac{v_{\gamma ph}}{\operatorname{Re} n_1} \qquad (7)$$

**(ii)** High $\gamma$ – **energy SCR- photons** with the polarizations perpendicular on decay plane Q, with the emission SCR-angle given by the

$$\cos\theta_{12} = v_{\gamma ph}.v_{2ph} \cong \frac{v_{2ph}}{v_1 \operatorname{Re} n_1} \leq 1 \qquad (8)$$

## Quantum SCR theory

The quantum theory of SČR is similar to the quantum theory of ČR [5] and here we present only some final results for the case of a transparent nondispersive medium. So, just as in the quantum CR-theory the same interaction Hamiltonian $H_{fi}$ with some modifications of source fields in medium can also describe the coherent $\gamma$-emission in all sectors (LE and HE). Then it is easy to see that the intensity of Super-Cerenkov radiation can be given in the following form [50]

$$\frac{dN}{d\omega dt} = \frac{\alpha Z^2}{v_1} \frac{1}{|n_1|^2 |n_2|^2 |n_\gamma|^2} \frac{k}{\omega}\frac{dk}{d\omega} S \cdot \Theta(1-\cos\theta_{SC})$$

where the spin factor S, for a two body electromagnetic "decay" of a spin ½ $^+$ - particle in medium is given by

$$S \equiv \frac{(E_1+M)(E_2+M)}{4E_1 E_2}\left[\frac{p_1^2}{E_1+M}+\frac{p_2^2}{E_2+M}+2\frac{(\vec{e}_k\cdot\vec{p}_1)(\vec{e}_k\cdot\vec{p}_2)-(\vec{e}_k\times\vec{p}_1)(\vec{e}_k\times\vec{p}_2)}{(E_1+M)(E_2+M)}\right]$$

Fig. 4b. A brief description of quantum theory of SCR-decay

## Quantum SCR theory (continuation)

Now, one can see that $\Theta(1-\cos\theta_{SC})$ Heaviside step function is 1 in two (or many) physical regions defined by the constraints: $\cos\theta_{2\gamma}=v_{2ph}(E_2)v_{\gamma ph}(\omega)<1$. The spin factor S in the above formulas is defined just as in the usual CR-quantum theory but with the particle's momentum $p_i$, i= 1,2, considered in medium. The vector $e_k$ is the photon polarization for a given photon momentum k. For a given k we choose two orthogonal photon spin polarization directions, corresponding to a polarization vector perpendicular and paralel to the SČR-decay plane Q.

$$\vec{e}_k \perp Q-\text{plane}$$
$$S^\perp \equiv \frac{(E_1+M)(E_2+M)}{4E_1 E_2}\left[\frac{\vec{p}_1}{E_1+M}-\frac{\vec{p}_2}{E_2+M}\right]^2$$

$$\vec{e}_k \parallel Q-\text{plane}$$
$$S^\parallel \equiv \frac{p_1 p_2}{E_1 E_2}\sin\theta_{1\gamma}\sin\theta_{2\gamma}$$

Fig. 4c. A brief description of the quantum theory of SCR-decay

Hence, the main signatures of the SCR-phenomenon are as follows:
- The SCR-effect in the low energy sector with the SCR-coherence condition (6) instead of CR-condition (1). So, the subthreshold CR-effects will be observed up to the SCR-threshold (7). The usual CR is a limiting process when $\operatorname{Re} n_1 \to 1$.
- The SCR-effect in the high energy sector (see Fig. 4a) have two main signatures, namely, the secondary SCR-effects responsible for anomalous CR-rings (see Fig. 3a), and possible secondary SCR-effects (produced by high-energy gamma} responsible for the concentric CR-rings.

Therefore, the problem of the experimental test of Super-Cerenkov coherence condition (6) is of great interest not only for the fundamental physics but also for practical applications to the particle detection. Such a test was performed by us [51] by using the experimental data of Debbe et al. [52] obtained at BNL with a ring imaging Cerenkov Detector (RICH). In the RICH detectors, particles pass through a radiator, and a spherical mirror focuses all photons emitted at angle $\theta_{SC}$ (see Fig. 5a) along the particle trajectory at the same radius $r_{SC}(p) = \frac{R}{2}\tan\theta_{SC}$ on the focal plane. Photon sensitive detectors placed at the focal plane detect the resulting ring images in the RICH detector. So, RICH-counters are used for identifying and tracking charged particles. Cerenkov rings formed on a focal surface of the RICH provide information about the velocity and the direction of a charged particle passing the radiator. The particle's velocities are related to the Cerenkov angle $\theta_{CR}$ or to the Super-Cerenkov $\theta_{SCR}$ by the relations (1) and (2), respectively. Hence, these angles are determined by measuring the radii of the rings detected with the RICH-detectors. In ref. [52] a $C_4F_{10}Ar(75:25)$ filled RICH-counter read out was used for measurement of the Cerenkov ring radii.

**Experimental tests of SČR predictions**

1. An experimental test of the SČR predictions can be obtained using RICH detectors. For this <u>accurate measurements of the SČR-ring radius are needed</u>
2. As is well known, in an usual RICH detector a spherical mirror focuses all photons emitted at a Čerenkov angle along the particle trajectory at the same radius on the focal plane. Photon sensitive detectors placed at the focal plane detect the resulting ring image.
3. Then, a fundamental test of SČR can be based on SČR-prediction on ring radii:

$$r_{SC}(p_X) = \frac{R}{2}\tan\theta_{SC}$$

Fig. 5a. Schematic description of the SCR fundamental tests based on the SCR-ring radii predictions

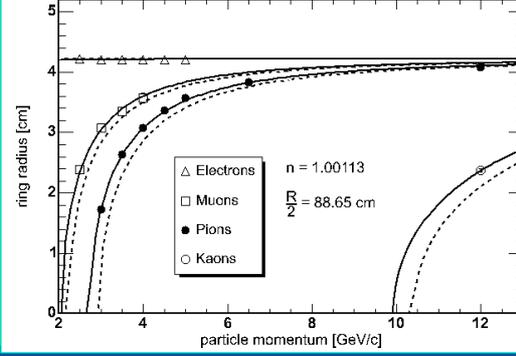

**Fig. 5b.** Experimental Čerenkov ring radii of the particles e, μ, π, K, obtained by Debbe et al. [52] with RICH detector, are compared with the theoretical *Super-Čerenkov prediction* (solid curve), and also with the *Čerenkov prediction* (dashed curves)

In Fig. 5b we presented the experimental values of the ring radii of electrons, muons, pions and kaons measured in the active area of this RICH-detector. The saturated light produced from electrons was a decisive fact to take in [52] an index of refraction $n_\gamma$ =1.00113 for the radiator material. The absolute values for excitation curves of electron, muon, pion and kaon, shown by dashed curves in Fig. 5b, was obtained by using this value of refractive index in formula: $r_{CR}(p) = \frac{R}{2} \tan\theta_{CR}$. The solid curves show the individual best fit of the experimental ring radii with SCR-predictions (6).

**Table 1**: The results of the best fits of the experimental
Čerenkov ring radii r(p) of Debee et al.[52] with SČR prediction: $r_{SCR}(p)$
given in Fig.5a,b,c for $n_\gamma = 1.00113$

| Particle | Number of data | $10^3 a^2 (GeV/c)^2$ | $\chi^2/n_{dof}$ |
|---|---|---|---|
| e | 6 | -0.081±0.101 | 0.468 |
| μ | 4 | 1.449 ±0.098 | 3.039 |
| π | 7 | 2.593 ±0.167 | 0.234 |
| K | 1 | 21.140 ±2.604 | - |
| All data | 18 | $\left[\frac{a}{m}\right]^2 = 0.1211 \pm 0.0053$ | 1.47 |

In the papers [50,51] we fitted all the 18 experimental data on the ring radii from [52] with our Super-Cerenkov prediction and we obtained the consistent result presented in Fig. 5c and Table 1. For other results on SCR classical and quantum theories, SCR-Yield, CR-Yields, and their differences which are maximum at Cherenkov thresholds, see our paper [51]

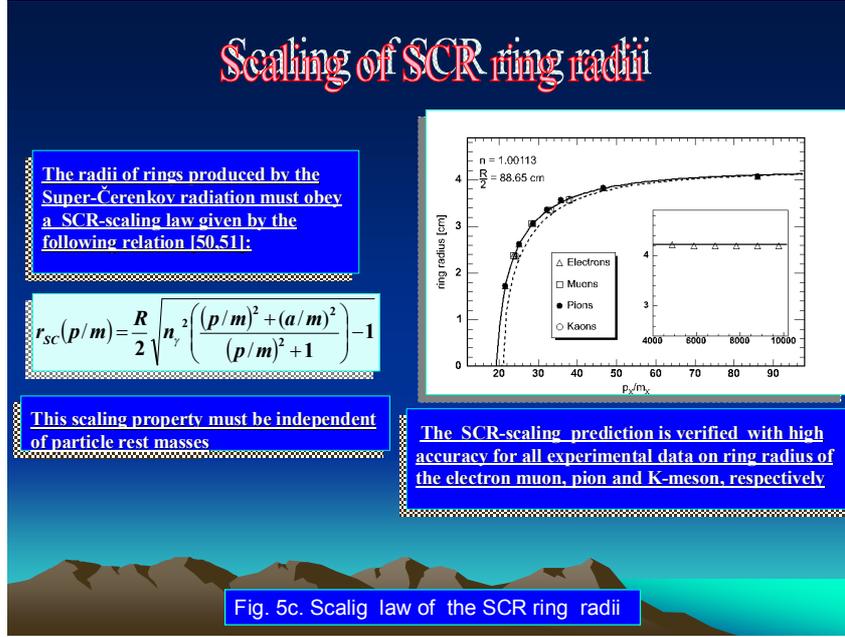

Fig. 5c. Scalig law of the SCR ring radii

### 4. Generalized SCR-Radiation:

Generalized Cerenkov-like effects based on four fundamental interactions was investigated and classified in our paper [35]. Then, it was shown that in order to obtain the spontaneous particle production via Cherenkov-like effects in a medium, three general conditions are to be fulfilled:

*(i-CR) The incident particle-source must be coupled to a specific radiation (electromagnetic, mesonic, fermionic-antifermionic, etc.) field ( RF);*

*(ii-CR) The propagation properties of the RF must be modified inside the medium in such way that the phase velocities of the RF-quanta in that medium are less than the light velocity in vacuum.*

(iii-CR) *The particle-source must be moving in the medium with a velocity v higher than phase velocity $v_{ph}$ of the RF in that medium.*

Then it is easy to see that three kinds of generalized Cherenkov-like effects can take place in nuclear media. Namely, we have (see Ref. [35])
I. The strong Cherenkov-like effects.
II. The electromagnetic Cherenkov-like effects.
III. The weak Cherenkov-like effects.
IV. Gravitational Cherenkov-like effects.

Now, in order to obtain the spontaneous particle production via generalized Super-Cherenkov-like radiation in medium, three general conditions are to be fulfilled:

(i-SCR) *The incident particle-source must be coupled to a specific radiation (electromagnetic, mesonic, fermionic-antifermionic, etc.) field ( RF);*

(ii-SCR) *The propagation properties of the RF (M-quanta) as well as those particle source must be modified inside the medium.*

(iii-SCR) *The particle-source (X) must be moving in the medium with a phase velocity*
$v_{Xph}(E_1)$ *such that the SCR-decay condition:*

$$\cos\theta_{MX} = v_{Mph}(\omega) \cdot v_{Xph}(E_1) \leq 1 \tag{8}$$

*is fulfilled*.

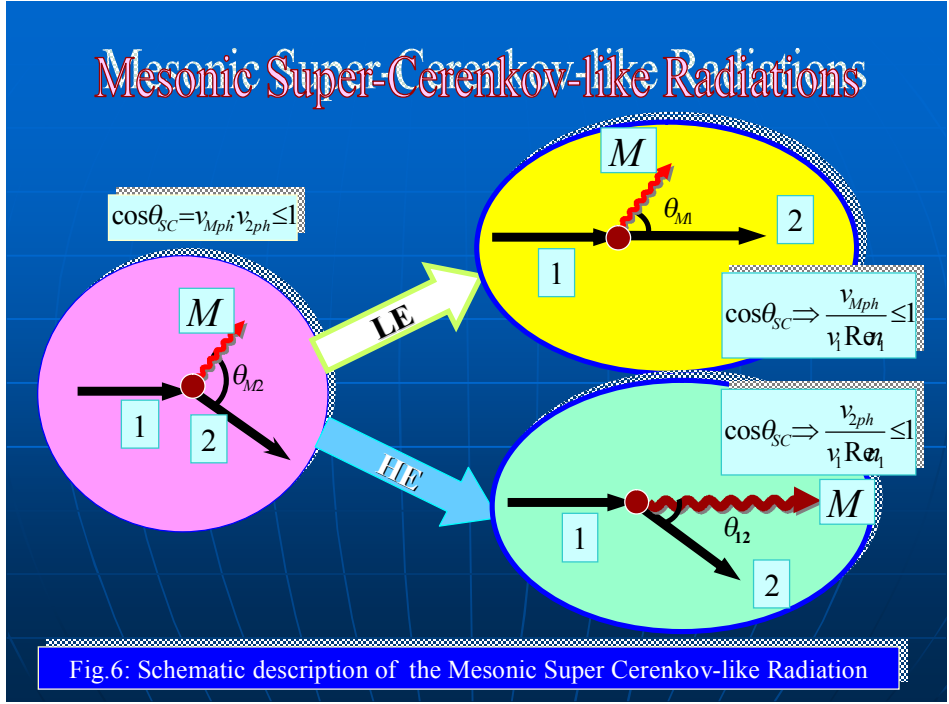

Fig.6: Schematic description of the Mesonic Super Cerenkov-like Radiation

As an example we consider here the possibility of a M-mesonic emission via mesonic Super-Cerenkov mechanism inside of a nuclear (or a hadronic) medium. For a schematic description of the generalized mesonic Super-Cerenkov-like radiation see Fig.6

Moreover, based on four fundamental interactions and on above above (i-SCR)-(iii-SCR) generalized conditions, we obtain the generalized SCR classification presented in Fig. 7.

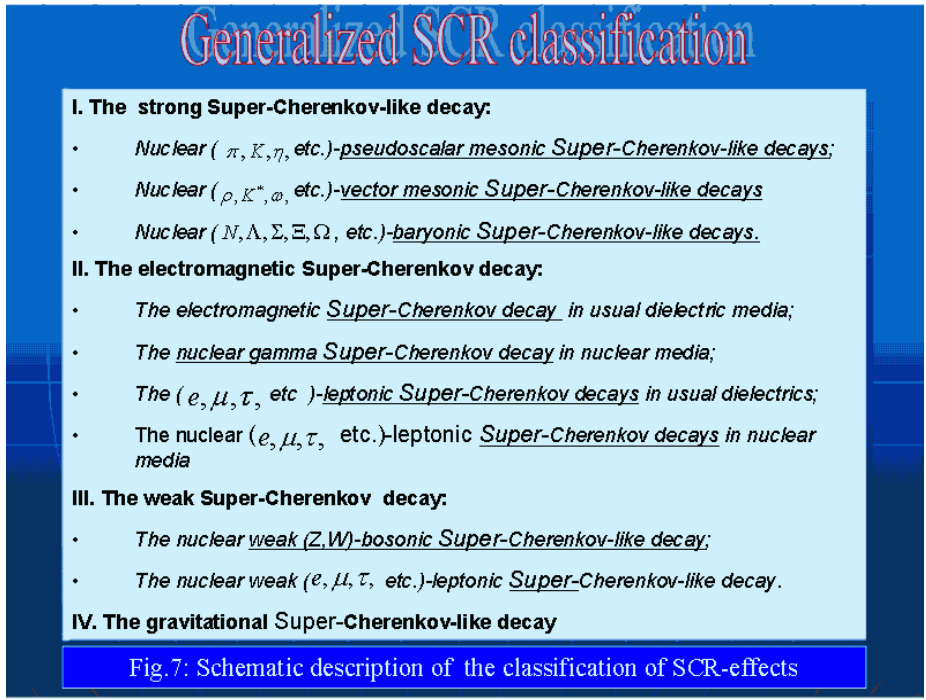

Fig.7: Schematic description of the classification of SCR-effects

## 5. The gluonic Super-Cerenkov-like radiation

According to the elementary particle classification (see Fig. 8) we can introduce not only $\gamma(gamma) - SCR$, but also some new concepts such as: g(gluonic)-SCR as well as $Z(bosonic) - SCR$ and $W(bosonic) - SCR$, elementary super-Cerenkov radiations.

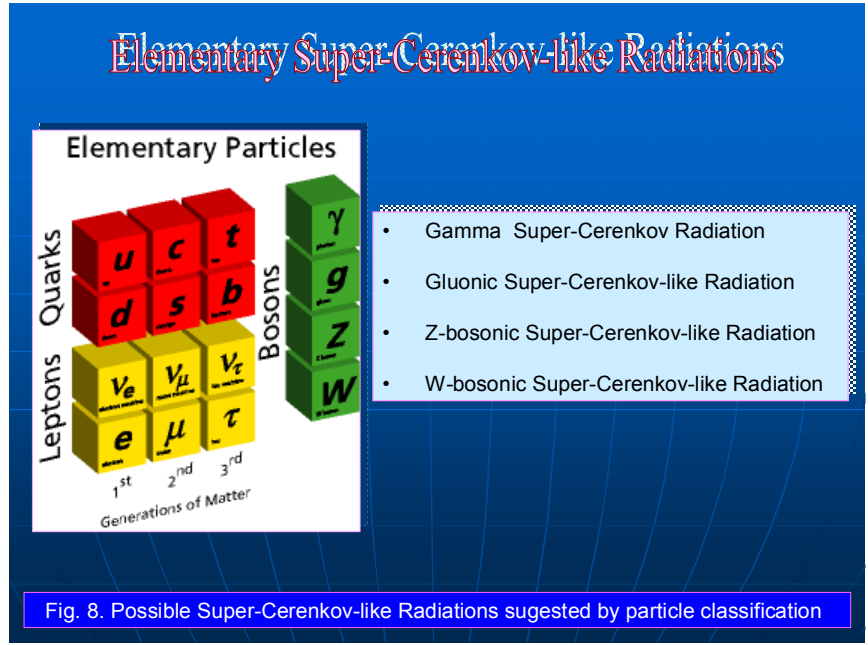

Fig. 8. Possible Super-Cerenkov-like Radiations sugested by particle classification

It is well know that that partons cannot be produced as part of a final state of a process. Partons are essentially quarks, which cannot be produced as free particles. However, distinct 'jets' of particles can be produced in the final state of deep inelastic scattering when the virtual boson interacts with a gluon in the proton to produce a quark-antiquark pair. As these partons move away from the interaction region the strong force becomes so strong that they cannot emerge as free particles. They fragment into jets of long-lived particles which are collimated along the direction of the original parton. According to this interpretation, it was speculated by many authors [25-26, 53-57]. that the partons which are moving in hadronic matter with the velocities higher than phase velocity of gluons can produce Cerenkov gluons analogous to Cerenkov photons. The notion of Cerenkov gluons was first introduced by Dremin [25-26] to explain some experimental cosmic data [58]. The difficulties are connected with the fact that the essential ingredients such as refractive index of gluons inside the hadronic medium cannot be predicted or tested experimentally with high accuracy as in the case of photons or mesons, baryons or leptons, inside of dielectrics, or even nuclear media [35]. So, the results in the domain of Cerenkov gluons remain mainly at speculative levels and a systematic theoretical approach on this problem seems to be far from to be constructed.

However, a systematic theoretical approach of the gluonic SCR-like effect can be developed in agreement with QCD and with the schematic description from Fig. 9. Therefore, let $v_{1ph}$, $v_{2ph}$ be the parton phase velocities in hadronic medium in the initial and final state in the SCR-decay from Fig.9, and $v_{gph}$ the gluon phase velocity in the same hadronic medium. As we see from Fig. 9, two kind of Super-Cerenkov-gluons (SCG) are predicted. These are as follows:

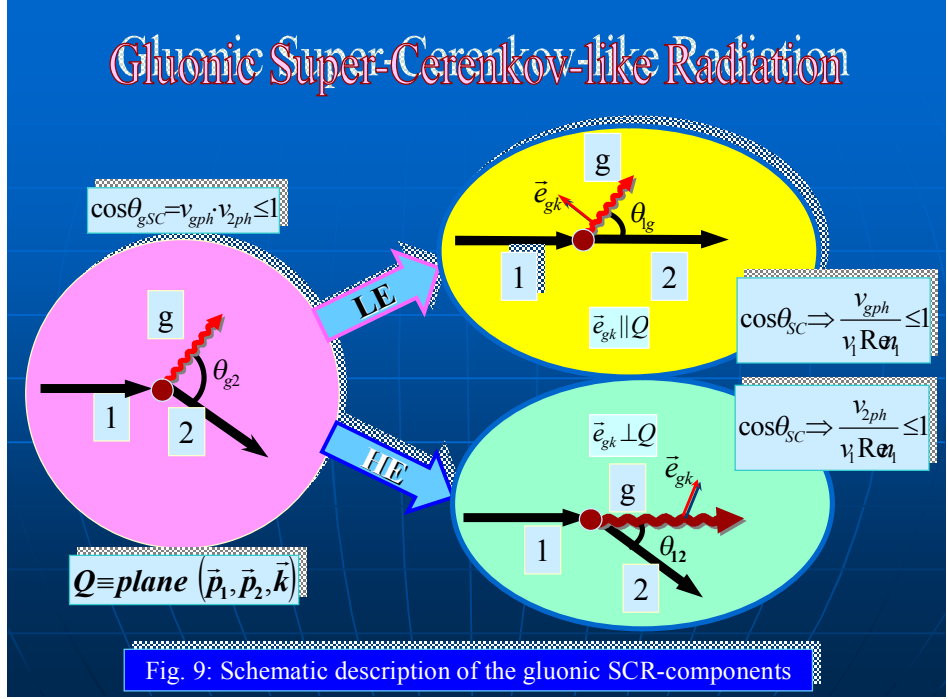

Fig. 9: Schematic description of the gluonic SCR-components

(i) <u>SCR-gluons of low energies</u> [or Cerenkov gluons] with the polarization in the decay Q-plane and with the emission SCR-angle given by

$$\cos\theta_{SCG} = v_{1ph} \cdot v_{gph} = \frac{v_{gph}}{v_1 \operatorname{Re} n_1} \leq 1, \qquad (9)$$

(ii) <u>SCR-gluons of high energies</u> with polarization perpendicular on the Q-plane, with the emission SCG-angle given by the

$$\cos\theta_{SCG} = v_{1ph} \cdot v_{gph} \cong \frac{v_{2ph}}{v_1 \operatorname{Re} n_1} \leq 1 \qquad (10)$$

where by $n_i(E_i)$, $n_g(\omega)$, i=1,2, are the partons and gluons refractive indices in hadronic medium which, in principle, can be determined according to the Foldy-Lax formula

$$n_X^2(E_X) = 1 + \frac{4\pi\rho}{E_X^2 - M_X^2} \cdot C(E_X) \cdot \bar{f}_{Xc \to Xc}(E_X, 0^o), \qquad (11)$$

where $E_X, M_X$ are the energy and rest mass of particle X (e. g. a parton, a gluon, etc.) in the hadronic medium, while **c**-are the scatterers centers (quarks, partons, etc); *$\rho$ -is the density of the scatterers centers in hadronic medium*; $\bar{f}_{Xc \to Xc}(E_X, 0^o)$-is the average ($Xc \to Xc$)-elastic scattering amplitudes in the forward direction; $C(E_X)$ – is the coherence factor of the hadronic medium. It is important to note that $C(E_X) = 1$ if the scatterer centers are randomly distributed.

Finally, we note that RHIC experiments [56,57] have shown two bump structure of the azimuthal distributions near the away-side jets. This can be interpreted as the signature of the SCR-gluon emission. However, the alternative interpretation as direct signature of the generalized mesonic, baryonic, leptonic, etc., SCR-effects (see the classification from Fig. 7), cannot be avoided. Of course more theoretical and experimental investigations are necessary to clarify the problems of the SCR-gluons emissions in hadronic media.

## 6. Summary and conclusions

In this paper the generalized Super-Cerenkov-like effects as new dual coherent particle production mechanisms are briefly presented. The main results and conclusions can be summarized as follows:
1. The first confirmation of the NPICR predictions [35] in Dubna experiments [37] is illustrated in Fig. 2a;
2. The description with high accuracy of the integrated cross section of the single meson production via mesonic Cerenkov-like effects in hadronic media (in the variant HMCR ) is reviewed in Fig. 2b;
3. The experimental results of Vodopianov et al., on the anomalous Cerenkov rings as one of the most important signature of the Super-Cerenkov radiation (SCR) are shown in Fig.3a;
4. The concept of Super-Cerenkov Radiation (SCR), as well as the SCR-signatures are schematically described in Fig. 4a.
5. The main results on the quantum theory of the SCR-phenomenon are briefly presented in Figs. 4b,c;
6. The results of the first experimental test of the super-coherence condition (2), obtained by using the experimental data from BNL [52], are presented in Fig. 5b and Table 1;
7. The scaling properties of the SCR-ring radii are illustrated in Fig.5c;
8. The concept of the Super-Cerenkov mesons, first introduced here, is schematically described in Fig.6.
9. A classification of the generalized SCR-effects based on four fundamental interactions is given in Fig.7;
10. The new concepts such as: SCR-gluons, SCR-W-bosons and SCR-Z-bosons, all three suggested by elementary particle classification, are introduced (see Fig. 8);
11. The gluonic Super-Cerenkov-like radiation, first introduced here, is schematically described in Fig. 9. The interpretation of some recent RHIC results as signature of the SCR-gluons is suggested.

Finally, that we believe that the results presented here are encouraging for more theoretical and experimental investigations which are necessary to clarify the problems of the coherent gluons emissions as gluonic Super-Cerenkov-like radiation in hadronic media.